# Data-Driven Prediction of Dielectric Anisotropy in Nematic Liquid Crystals

Charles Parton-Barr[1] and Richard. J. Mandle* [1,2]

[1] School of Chemistry, University of Leeds, Leeds, UK, LS2 9JT
[2] School of Physics and Astronomy, University of Leeds, Leeds, UK, LS2 9JT

*r.mandle@leeds.ac.uk

## Abstract

We curate a large-scale dataset of low frequency dielectric anisotropy values for low molecular weight liquid crystals. Using this dataset, we demonstrate that supervised machine-learning models can predict dielectric anisotropy with substantially improved accuracy (RMSE 2.6) compared to estimates obtained from the Maier-Meier relations using molecular properties from both the widely used semiempirical AM1 method (RMSE 9.7) and the modern r2scan-3c composite method (RMSE 11.2). Realising the potential of machine learning techniques for liquid crystalline materials requires carefully curated data to be accessible, and on this basis we propose a simple and standard template for reporting data.

## Introduction

Nematic liquid crystals are widely exploited in display technology, making the ability to accurately predict their macroscopic properties from molecular structure a central challenge in materials design. In contrast to other functional and soft materials, liquid crystalline materials must be synthesised in high chemical purity and are (typically) characterised as bulk materials in the condensed anisotropic state, rather than in dilute (isotropic) solution. As a result, experimental screening of materials is intrinsically low-throughput, time-consuming, and resource intensive. This creates a particularly acute need for accurate predictive structure-property relationships, especially tools and methods capable of linking molecular scale properties to measurable bulk properties. Of the many properties of nematic liquid crystals, the dielectric anisotropy (Δε) is of particular importance, governing electro-optic response, switching behaviour, and device operating voltages. Dielectric anisotropy is often calculated according to the Maier-Meier relationship, effectively an extension of Onsager theory to anisotropic systems with long range orientational order, given in eq (1):

$$\Delta\varepsilon = \frac{NFh}{\varepsilon_0}\left\{\Delta\alpha - \frac{F\mu_{eff}^2}{2k_BT}(1-(3cos^2\beta))\right\}S \qquad (1)$$

Where $\mu_{eff}^2 = \sqrt{g.\mu^2}$, $\mu$ is the molecular electric dipole moment, $g$ is the Kirkwood g factor, $\beta$ is the angle between the electric dipole moment and the molecular long axis, $\Delta\alpha$ is the anisotropic polarisability, $N$ is the number density, $F$ is the reaction field factor, and $h$ is the cavitation factor which are defined according to:

$$F = \frac{1}{1-(\varepsilon - \frac{1}{(2\pi\varepsilon_0 a^3).(2\varepsilon+1)})} \qquad (2)$$

$$h = \frac{3\varepsilon}{(2\varepsilon+1)} \quad (3)$$

The AM1 method has been successfully used to calculate dielectric anisotropy from computed dipole moments and polarisability tensors. [1-5] Despite this empirical success, the AM1 method is known to significantly underestimate polarizability; [6, 7] its apparent success when used with the Maier-Meier relations demonstrating that numerical agreement does not imply physical fidelity. More generally, while the Maier-Meier relation provides a mean-field link between computable (or measurable) molecular properties and macroscopic dielectric response, its underlying assumptions (single molecule response, weak treatment of intermolecular correlations) can restrict its quantitative accuracy.

Hybrid DFT methods have been used with Maier-Meier relations with some success, [8, 9] yet the application of composite methods – which combine high chemical accuracy with low computational cost - remain unexplored. At the same time, the growing availability of chemically diverse datasets, together with rapid progress in (un)supervised molecular machine learning, raises the question of whether dielectric anisotropy can be predicted directly from molecular structure without reliance on explicit mean-field models.

Here we address this question using supervised machine-learning models trained to predict $\Delta\epsilon$ of nematic liquid-crystalline materials. In the absence of any public dataset of experimental $\Delta\epsilon$ data for individual molecules exists, we curated our own for this purpose, drawing from both academic and patent literature. Three machine-learning model architectures were trained: multilayer perceptrons (MLP), graph neural networks (GNNs) and XGBoost Regressors. The predictive performance of these models was benchmarked against dielectric anisotropies estimated using the Maier–Meier relation, employing both a widely used semi-empirical electronic-structure method (AM1) and a modern, low-cost composite density-functional approach (r²SCAN-3c). This comparison enables a direct assessment of data-driven and physics-based approaches for predicting dielectric response across chemically diverse liquid-crystal systems. Finally, SHAP analysis is performed on the best performing fingerprint model to evaluate the relationship between descriptor values and $\Delta\varepsilon$ prediction.

## Method

### Dataset Curation

We curated a dataset of low frequency dielectric anisotropy data for single component low molecular weight liquid crystals. The dataset covers around 3500 unique molecular structures, drawn from both journal [10-129] and patent literature. [130-500]

The dataset includes data measured for pure materials and that obtained by extrapolation from mixtures. The dataset does not include data for mixed systems (e.g. unextrapolated mixtures, mixtures of isomers, low purity materials). Where the same material was reported in multiple papers and/or patents, we use values from the earliest publication. Molecular structures were transcribed using a text-based notation system and converted to SMILES. The raw dataset was handled using RDkit, [501] allowing us to filter inadvertent duplicates due to human error (e.g. where two identical molecules have different "raw" SMILES due to differing ring indices).

**Electronic Structure Calculations**

We converted SMILES strings into 3D coordinates using the ETKDGv3 conformer generator [502] in RDKIT (Release_2025.09.4). We then performed geometry optimisation and calculation of dipole moment and polarisability tensors using the AM1 semi-empirical method (Gaussian G16 [503]) and the r2scan-3c composite method [504] (Orca 6.1[505]).

**Dataset Informatics**

We computed descriptors (Molecular weight, fraction Csp³) using the Descriptors module in RDKit. Uniform manifold approximation and projection (UMAP) was performed with the UMAP-learn package [506] using Mordred fingerprints. Prior to embedding, descriptors containing non-finite values (NaN or ∞) were removed, and molecules with remaining non-finite entries were excluded. The cleaned descriptor matrix was embedded into two dimensions using UMAP with n_neighbors = 20, min_dist = 0.1, and a fixed random seed to ensure reproducibility.

**Machine Learning**

Molecular fingerprints (1024 bit) were generated using the *skfp.fingerprints* Python package.[507] For three-dimensional fingerprints we generated low-energy conformers from the SMILES string using the ETKDGv3 algorithm in RDKIT. [502] Machine learning models were trained using supervised learning framework using PyTorch. [508] The molecular fingerprints were used as the input, and the values of dielectric anisotropy as the target. The dataset was randomly divided into test/training sets using a 80:20 split. Model training was performed for 500 epochs (with early stopping) using the Adam optimiser with a learning rate of $1 \times 10^{-3}$ and a weight decay of $1 \times 10^{-4}$. A batch size of 64 was employed throughout. Model performance was evaluated using 5-fold cross-validation on the training data. Predictive accuracy was assessed on the held-out test set. All fingerprint neural networks shared the same topology (Figure 1). Hyperparameter choice was made empirically based on the need to balance training stability and predictive performance.

Fingerprint-based multilayer perceptrons (MLPs) operate on fixed-length fingerprint vectors which encode molecular information. The MLPs learn to map to a target property through the transformations of successive hidden linear layers. Graph neural networks (GNNs) instead are passed molecular graph representations and learn atomic-level representations through message-passing convolutional layers. These atomic-level representations are then pooled into a molecular-level prediction. As a non-neural-network comparison, the XGBoost regressor builds an ensemble of decision trees, each trained to correct the residual errors of the previous tree.

Molecular graph representations were produced using previously published code. [509], and the models were trained according to the hyperparameters described in the ESI . An XGBoost [510] regressor model was trained in a manner as to replicate the MLP training process. It was parameterised with 500 estimators, a max depth of 10, a learning rate of 0.01.

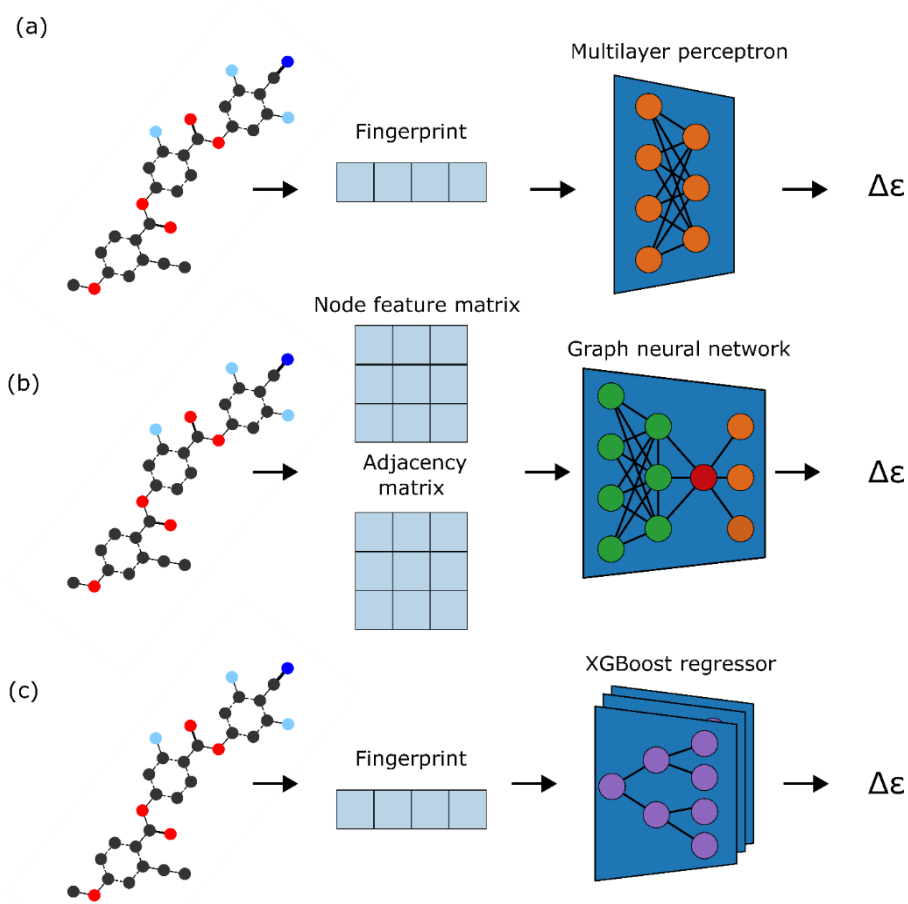

**Figure 1:** Schematic representation of the machine learning architectures used in this work. (a) illustrates the multilayer perceptron that transforms a molecular fingerprint through successive hidden layers to predict $\Delta\varepsilon$. (b) Shows how a graph neural network where the molecule is represented by a node feature matrix and an adjacency matrix. The information of neighbouring nodes is aggregated through message passing layers (green), then pooled to a graph level representation (red). A scalar prediction is then produced through the linear transformation layers (orange). (c) depicts the XGBoost regressor in which an ensemble of decision trees predicts $\Delta\varepsilon$ via sequential error correction.

## Results

### Dataset Analysis

Although the curated dataset is a broad cross-section of reported materials and design strategies, it is not (nor could it be) an exhaustive list of all possible materials nor all reported materials. By necessity, it the present dataset is shaped by experimental accessibility, by academic interest, by commercial relevance, and by disclosure. The iterative optimisation around certain building blocks (e.g. fluorinated aromatics) leads to clustering. The dataset comprises ~43% materials with negative dielectric anisotropy (N=1542, average $\Delta\epsilon$= -6.7) and ~57% with positive dielectric anisotropy (N=2065, average $\Delta\epsilon$ = +15.75). The average molecular weight across the dataset was 416 with a standard deviation of 118 (Figure 2b).

Molecules with large negative or positive values of $\Delta\epsilon$ typically feature a relatively modest fraction of saturated hydrocarbon; as the fraction of $Csp^3$ tends towards one the values of dielectric anisotropy cluster in the range 0-10. The dominant feature here is variance rather than trend.

The chemical space within the dataset is highly structured giving the UMAP projection a filament-like appearance; positive and negative materials are broadly interspersed throughout this space, although in some regions a clear bias to one sign of $\Delta\epsilon$ can be seen (Fig 2d). However, as UMAP is a non-linear embedding optimised for *local* neighbourhood preservation, distances and apparent clusters within the projection should not be over interpreted.

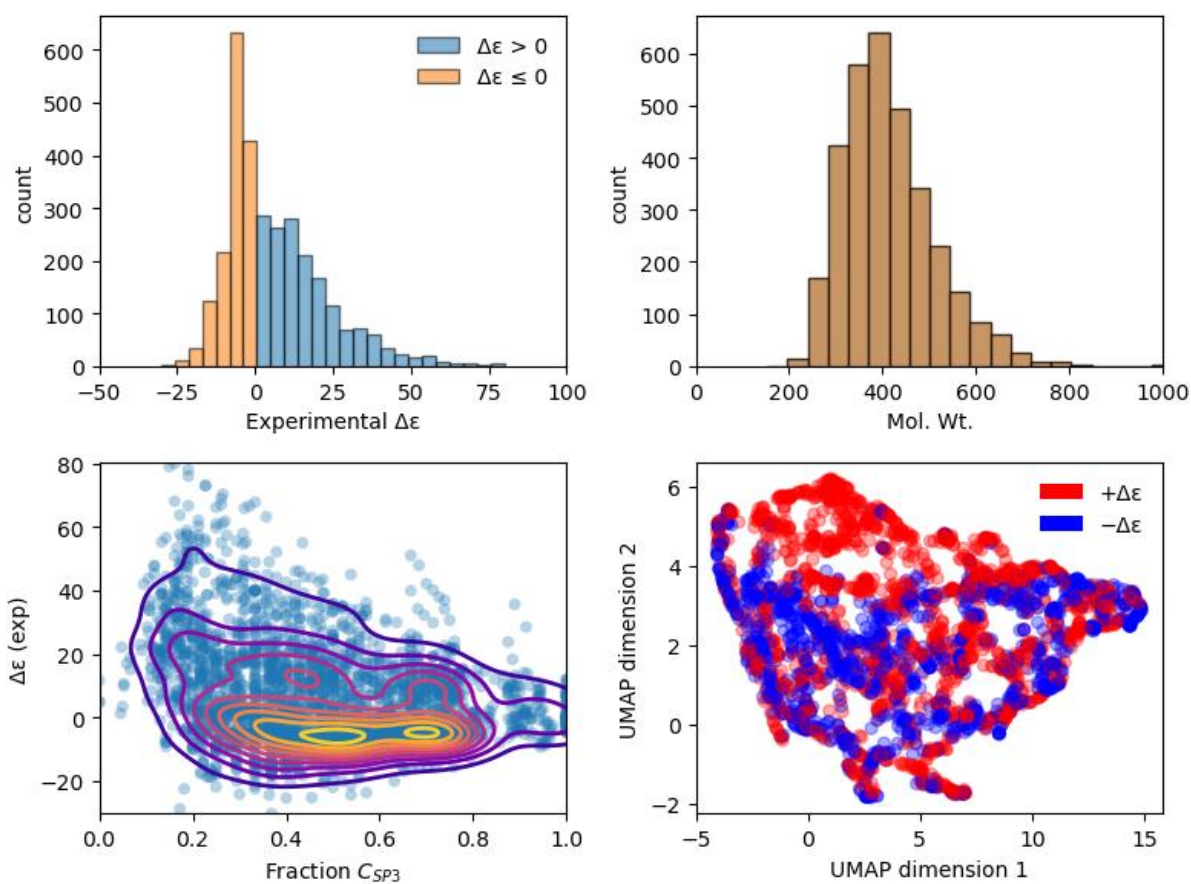

**Figure 2:** Analysis of the curated dataset. (a) Histogram plot of the reported values of dielectric anisotropy, coloured according to sign. (b) Histogram plot of molecular weight for molecules within the curated dataset. (c) Scatter plot of experimental dielectric anisotropy versus the faction of the molecule comprising *sp³* hybridized carbon, where solid contours represent the estimate point density in order to highlight the regions of highest data concentration. (d) UMAP projection of Mordred molecular fingerprint descriptors, with points coloured according to the sign of dielectric anisotropy.

**Electronic Structure Calculations**

Initial 3D geometries for dataset entries were generated from SMILES strings using the ETKDGv3 conformer embedding method in RDKIT. Where the conformer embedding failed (typically due to a lack of parameters, e.g. for hypervalent sulphur in $SF_5$) the geometry was manually generated. Electronic structure calculations were performed using either the AM1 semiempirical method or the r2scan-3c composite method. For each molecule, and with each method, we perform geometry optimisation followed by a frequency calculation to confirm the absence of imaginary frequencies. We then compute the dipole vector and polarisability tensor. Using the Maier-Meier relation with the obtained dipole and polarisability data with $S = 0.65$, $\rho = 1.00$ g cm$^3$, $T = 298$ K, and $g = 1$, we obtain the dielectric anisotropy.

In contrast to earlier use of the AM1 method for calculating dielectric anisotropy, [1-5] the dataset used here is orders of magnitude larger. However we find AM1 actually performs rather poorly: the root mean square error (RMSE) is 9.66 which corresponds to approximately 0.61 $\sigma$ of the experimental distribution ($r^2 = 0.632$). While this is a moderate degree of correlation, the magnitude of error and presence of frequent sign errors (*vide infra*) limit the qualitative predictive utility of the AM1 method in a workflow for predicting dielectric anisotropy.

Modern composite DFT methods such as r2scan-3c offer high chemical accuracy with low computational cost. That said, for our dataset we find an RMSE of 11.2 with the r2scan-3c method, corresponding to approximately 0.71 $\sigma$ of the experimental $\Delta\epsilon$ distribution ($r^2 = 0.500$). This performance is notably worse than that of AM1, indicating that increased accuracy of the electronic structure calculation does not translate to an improved predictive accuracy for $\Delta\epsilon$ *via* the Maier-Meier relation.

Both AM1 and r2scan-3c have significant "sign errors", where a material with negative $\Delta\epsilon$ is calculated to be positive (AM1: 10%, 297 entries; r2scan-3c: 13%, 371 entries) or a material with positive $\Delta\epsilon$ is calculated to be negative (both 3% of all entries). The poor performance of both methods presumably reflects the non-ideal use of the molecular "long axis" for orientation, that the calculations are performed on an isolated molecule and extrapolated to the bulk, the neglect of conformers, variations in $\rho$, $S$, $g$ and so on.

Considering multiple conformers may be a viable route to improving the accuracy of the calculated dielectric anisotropy. The computational cost of this is not negligible however; the optimisation and property calculations with r2scan-3c for the dataset took over 2700 CPU hours on a modern HPC system (Aire HPC at the University of Leeds). A back of the envelope calculation (assume 12 CPU hours to generate and optimise the conformers (using ORCA-GOAT), and consider just 10 conformers per molecule) would increase this to ~65800 CPU hours (around 7.5 CPU years), which is not insignificant.

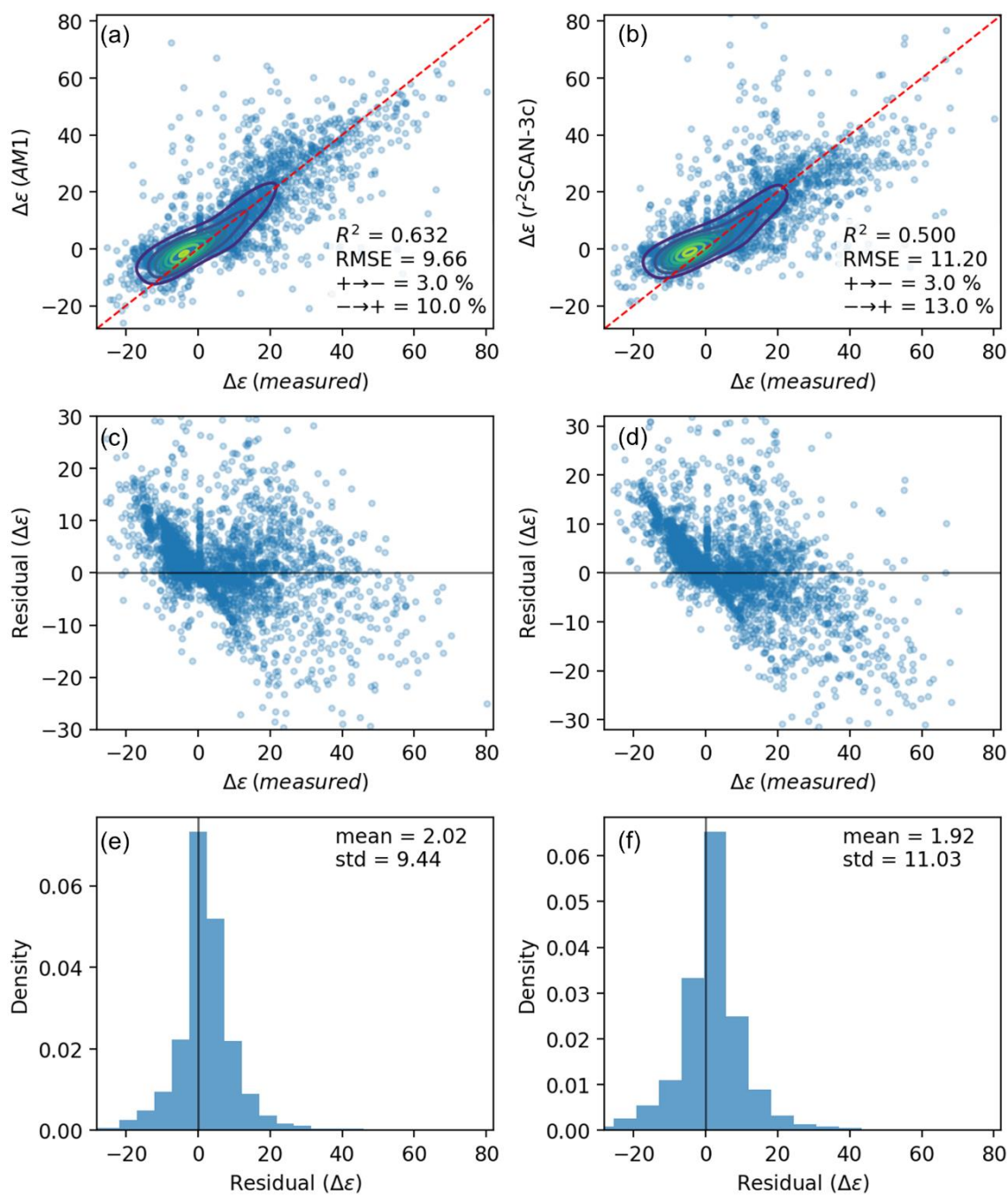

**Figure 3:** Plot of calculated (y) vs actual (x) dielectric anisotropy using geometry and properties from: (a) the semi-empirical AM1 method ($RMSE = 9.66$; $r^2 = 0.632$, and (b) the r2scan-3c composite method ($RMSE = 11.20$; $r^2 = 0.500$). Kernel density estimate (KDE) contours represent the estimated point density and so highlight the regions of highest data concentration. The dashed line represents perfect agreement (i.e. $x = y$). Both methods display significant sign inversions, as indicated by $+\rightarrow-$ (experimental $-\Delta\epsilon$, predicted $+\Delta\epsilon$) and $-\rightarrow+$ (experimental $+\Delta\epsilon$, predicted $-\Delta\epsilon$). Plots of residual vs actual $\Delta\epsilon$ for AM1 (c) and r2scan-3c (d). Histogram plot of residuals for AM1 (e) and r2scan-3c (f) with mean value and standard deviation labelled.

Plotting the residual $\Delta\epsilon$ ($\Delta\epsilon_{calc} - \Delta\epsilon_{meas}$) versus the experimental value (Figure 3c/d) reveals a clear negative slope, especially pronounced for r2scan-3c; in other words, an anticorrelation between prediction error and magnitude of the experimental value. This is particularly pronounced for r2scan-3c, and indicates a systematic deviation from ideal scaling, with increasingly strong underestimation at large positive and negative $\Delta\epsilon$.

The residual distributions (Figure 3e/f) are approximately Gaussian, but are shifted away from zero and with moderate-to-large variance (standard deviation). The combination of non-zero mean and significant variance reflects the presence of systematic error in the results and significant uncertainty across the entire dataset (as opposed to a few problematic structures).

Taken together, these diagnostics demonstrate that while electronic structure methods capture qualitative trends in dielectric anisotropy, their errors are structured rather than random. The presence of systematic anticorrelation in residuals indicate that factors governing the magnitude of $\Delta\epsilon$ are not fully captured by the models, and this motivates the use of data-driven approaches which we now turn to.

**Supervised Machine Learning**

Supervised machine learning is well suited to learning structure-property relationships from these representations in cases where no simple analytical model exists. This is particularly relevant for dielectric anisotropy, which emerges from properties across a range of length scales from the molecule (polarity, polarisability, electronic structure etc) to the bulk (e.g. order parameter, density).

Molecular fingerprints provide a numerical representation of chemical structure, encoding atomic connectivity, functional groups, and, in some cases, physicochemical descriptors. Such representations are particularly well suited to statistical learning, as they map chemically diverse molecules onto a common, fixed-length feature space. In this work, multiple fingerprint representations were generated for each molecule, each of these being a 1024-bit vector, enabling direct comparison and regression across the full dataset.

Molecular graphs are a representation in which atoms are treated as "nodes" and bonds as "edges", preserving the topology of the molecule without reliance on predefined descriptors. Unlike fingerprints (which are fixed-length features), graph representations retain information about relations between atoms; they are well suited to statistical learning and preserving chemically meaningful features that derive from the underlying molecular connectivity and topology.

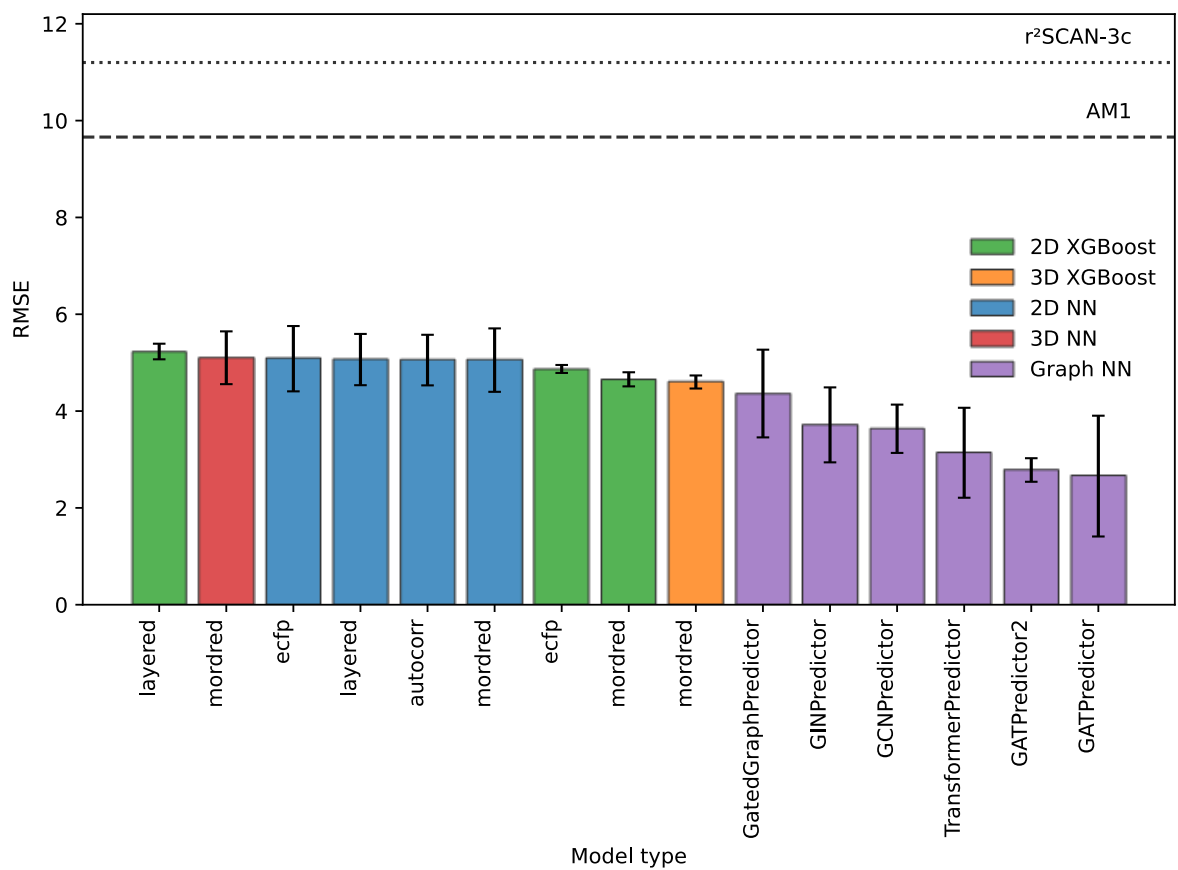

**Figure 4:** Performance of selected supervised machine-learning models for predicting dielectric anisotropy ($\Delta\varepsilon$) to compare a) molecular fingerprint vs graph representations and b) neural networks vs. XGBoost regressors. Bars show the root-mean-square error (RMSE) on the independent hold out test set for each model type; error bars show the standard deviation of the RMSE across cross-validation folds, arising from the repeated data partitioning during model training. The bar colours indicate the input used to train the network (blue = 2D fingerprint, red = 3D fingerprint, purple = molecular graph, green = 2D XGBoost, orange = 3D XGBoost). Horizontal dashed lines indicate the RMSE obtained for the same dataset when computing using the Maier-Meier relations with molecular properties as computed with the AM1 semi-empirical and r2scan-3c composite methods (*vide supra*). Additional examples of less performant networks are given in the ESI to this article, Figure S7 (fingerprint based) and Figure S8 (XGBoost regressors).

Three supervised machine learning approaches were employed to predict $\Delta\varepsilon$ values. The MLP and the XGBoost models were trained on fingerprint representations, whereas the GNN models were trained on molecular graphs. Across all methods, test-set RMSE values ranged from 2.6 to 20.0. The *erg* and *usr* fingerprint types performed worse than calculations via AM1/r2scan-3c and using the Maier-Meier relations (both having RMSE ≳ 10). We found 2D fingerprints to outperform those using 3D geometries, even for the same fingerprint type (e.g. *AtomPair*, *Mordred*). Out of all fingerprints tested, 2D Mordred gave the lowest RMSE for the independent hold-out test set (4.8) although others (*Autocorr*, *Layered*, *ECFP*, *Mordred-3D)* had only fractionally higher RMSE. Our non-neural network comparison,

XGBoost shows a slight outperformance of the MLP, with Mordred-3D, Mordred-2D and ECFP achieving RMSE values of 4.6, 4.7 and 4.9 respectively. The GNNs outperformed the fingerprint models in all cases, with the GATPredictor GNN achieving an RMSE of 2.6, again an improvement over the Maier-Meier relations

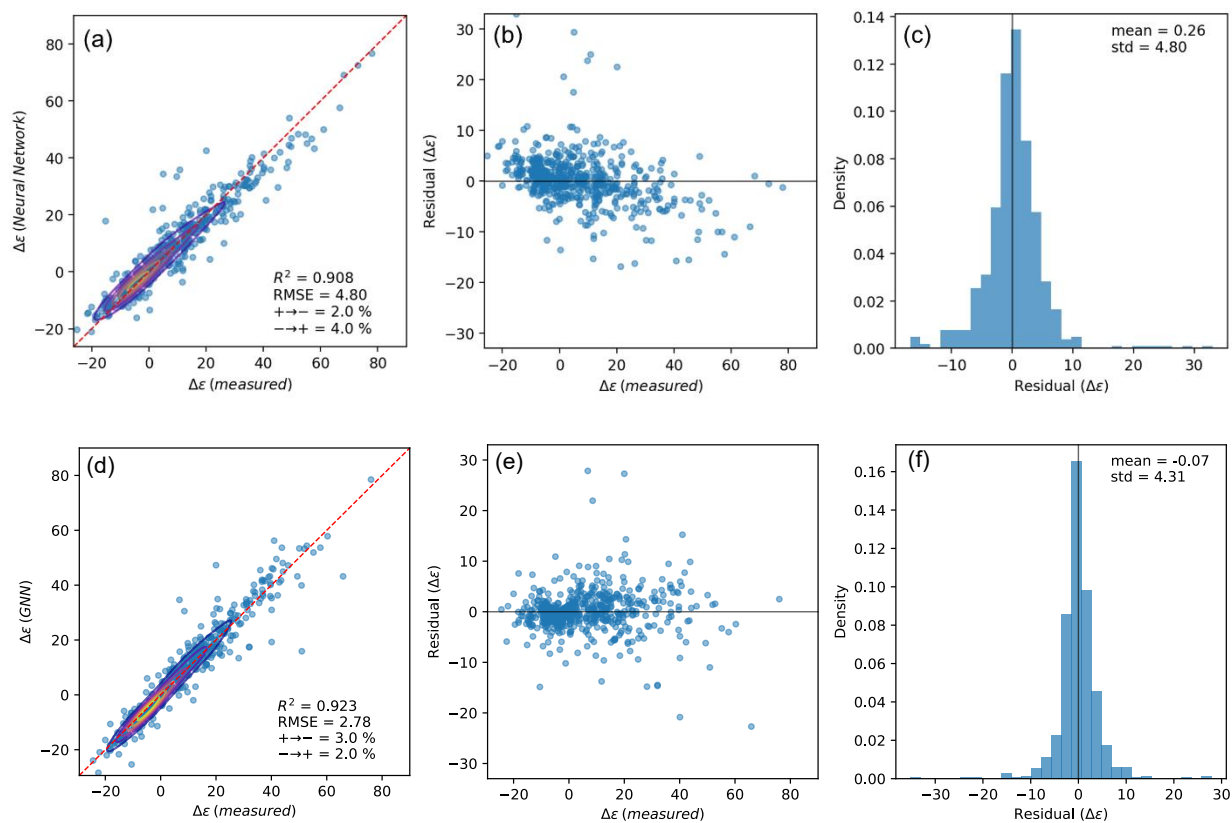

**Figure 5:** Fingerprint representation (top row) compared with graph representation (bottom row). Assessing the performance of the machine-learning model. Scatter plot of predicted $\Delta\epsilon$ versus experimental for the neural network trained against Mordred fingerprints (a) and the GATPredictor2 model (d). Kernel density estimate (KDE) contours are overlaid to highlight the distribution of data density, together with the parity line (dashed). The coefficient of determination ($R^2$) and root-mean-square error (RMSE) are indicated. (b,e) Residuals (predicted vs measured $\Delta\epsilon$) plotted as a function of the experimental value, showing the absence of systematic bias across the range of dielectric anisotropies. (c, f) Distribution of prediction errors for the independent hold-out test set, illustrating the spread and near-zero mean of the residuals.

The parity plot for the best-performing fingerprint model (2D Mordred Fingerprints, Figure 5a) compares the predicted versus experimental dielectric anisotropy. The model reproduces the experimental values with $r^2$ of 0.91 and RMSE of 4.8, indicating that much of the variance in $\Delta\epsilon$ is captured by the learned representation. By these metrics, the network is far more accurate than the Maier-Meier relation utilising AM1 or r2scan-3c. Predictions are well distributed about the $x = y$ line across the full range of experimentally observed $\Delta\epsilon$ values with no evidence of systematic compression at high magnitudes. Encouragingly, the percentage of sign errors is far lower. The corresponding residuals (Figure 5b) have a much weaker tendency to underestimation at large positive or negative $\Delta\epsilon$. Importantly, no discrete regimes or discontinuities are observed which would indicate fingerprint-specific artefacts, if present. The histogram of residuals (Figure 5c) is Gaussian (to a first approximation) and centred close to zero (mean = 0.26, standard deviation = 4.80) which confirms the absence of systematic bias in the model.

Although the GATPredictor architecture achieved the lowest test-set RMSE, figure 5 compares the Mordred-2D fingerprint model with GATPredictor2 results as this model achieved a lower standard deviation across its validation predictions (1.24 vs 0.24) which indicates a more stable model. GATPredictor2 achieves $r^2$ and RMSE values of 0.923 and 2.78 respectively. The parity plot (Figure 5d) shows a tighter clustering about the y = x line and the residuals (Figure 5e) present less of a break down in predictions at higher or lower $\Delta\varepsilon$ values. This is quantified in Figure 5f where the residual distribution is calculated to possess a mean of -0.07 and a standard deviation of 4.31 Together, these results demonstrate that both molecular fingerprints and graph-based approaches can learn the structural property relationships that govern dielectric anisotropy values in a chemically diverse dataset, with GNNs possessing stronger performance metrics.

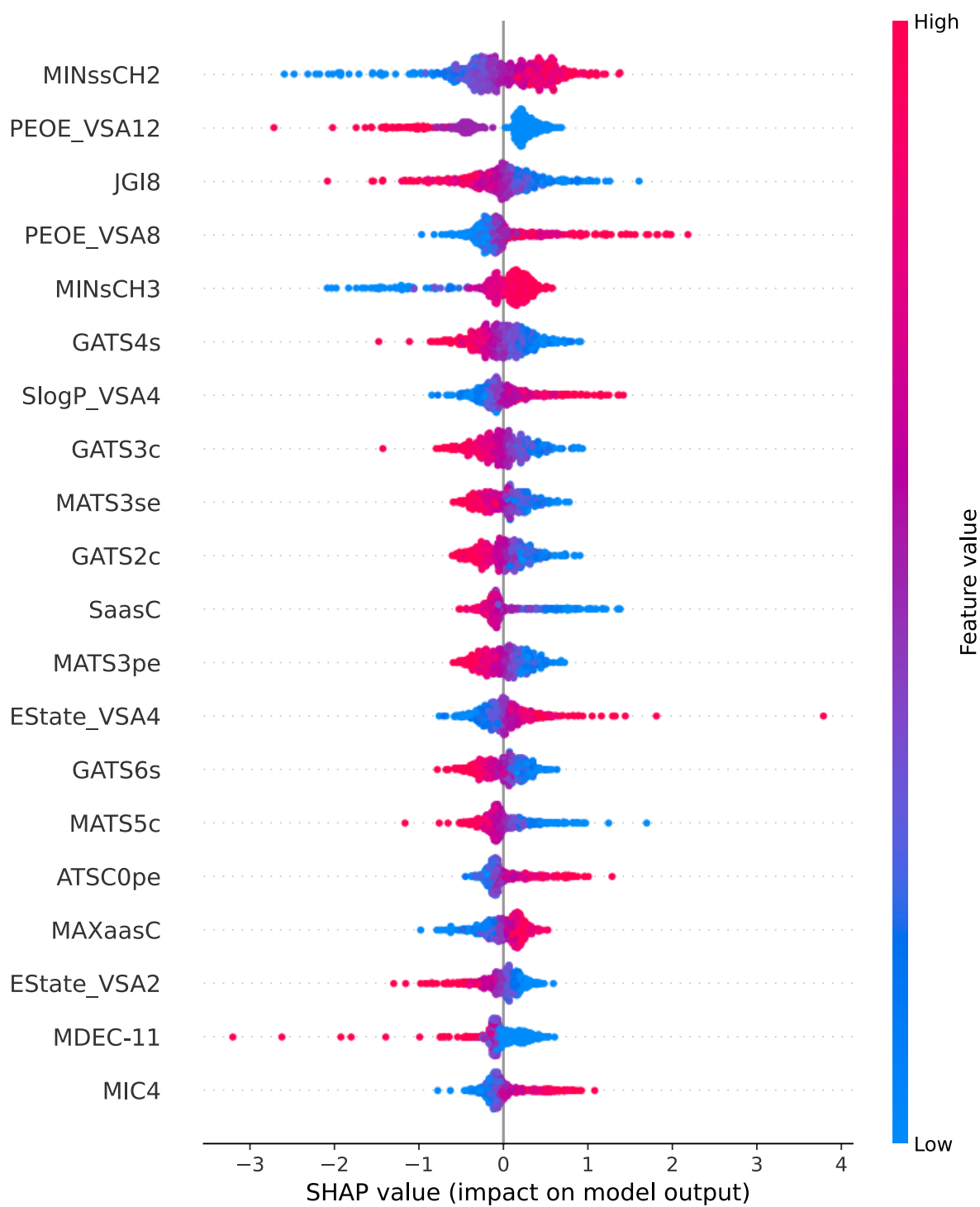

Figure 6. Calculated *SHapley Additive exPlanations* (SHAP) values for the best performing fingerprint neural network model. Descriptors within the Mordred fingerprint are ranked from most important (top) to least (bottom). Each point represents a unique molecule within the dataset. The colourscale reflects the normalised value of the descriptor value for each molecule. The position on the x-scale reflects the impact on the prediction of $\Delta\epsilon$.

Finally, we use SHAP values to explore interrogate the trained fingerprint model and identify which molecular descriptors most strongly influence the predicted value of $\Delta\epsilon$. SHAP values quantify the contribution of each descriptor to an individual prediction relative to the global model baseline. Positive SHAP values indicate features that increase the predicted dielectric anisotropy, whereas negative values correspond to descriptors that suppress it. Figure 6 presents a summary plot in which each point represents a single molecule, positioned according to the magnitude and direction of the descriptor's impact. Points are coloured by the normalised descriptor value, allowing the relationship between feature magnitude and model response to be visualised directly.

In this analysis we identify a set of high-impact descriptors spanning multiple physicochemically meaningful families. Prominent among these were topological autocorrelation metrics such as GATS4s, GATS3C, MATS5c, and ATSC0pe, which encode how electronic properties are distributed across the molecular graph. Surface-area descriptors partitioned by charge and lipophilicity (including EOE_VSA12, EState_VSA4, EState_VSA2, and SLOGP_VSA4) were likewise strongly represented, indicating that the exposure and spatial segregation of polar surface is important in governing $\Delta\epsilon$.

Extremal electro-topological descriptors (MINssCH2, MINsCH3, MAXaasC) and graph-based invariant descriptors such as MIC4 and MDEC-11 highlight the importance of electronically differentiated atomic environments and overall molecular topology. While individual descriptors do not map uniquely onto specific chemical fragments, the convergence of these independent feature classes on electrostatic distribution provides a physically coherent interpretation of the model. Collectively, the results indicate that dielectric anisotropy is governed primarily by the spatial organisation of electronic character rather than dipole magnitude alone, which we feel aligns with common intuitive understanding of this property.

Machine learning can only be successfully deployed where curated data is available. In the present case, this required careful curation of a large scale dataset spanning many journal articles and patents, coupled with transcription of chemical information into machine readable format. We would strongly encourage the inclusion of "machine readable" data in the ESI of future papers, and we give an example of this in the ESI to this article using examples from the dataset in this work. This need is especially acute for chemical structures; it is commonplace to use "X/Y/Z/R" groups and denote changes to structure in tabular form, and extracting the requisite information from data presented in this way is rather time consuming.

**Conclusions**

While physics-based models such as the Maier–Meier relation provide invaluable conceptual insight, their predictive performance is ultimately constrained by both their underlying assumptions and the accuracy with which molecular properties can be computed. We show these limitations to become particularly acute when such approaches are applied across large, chemically diverse datasets. In contrast, the machine-learning models reported here achieve substantially improved predictive accuracy without explicit recourse to mean-field assumptions. Notably, the most accurate predictions are obtained without explicit modelling of intermolecular correlations or nematic order. While such omissions are conceptually unsatisfying, even alarming, they do not diminish predictive performance – however, from a purely materials-design perspective, predictive accuracy is the most relevant currency.

**Acknowledgements**

RJM acknowledges funding from UKRI via a Future Leaders Fellowship, grant no. MR/W006391/1, funding from the University of Leeds via a University Academic Fellowship, and support from Merck KGaA.

Supplementary information for:

**Data-Driven Prediction of Dielectric Anisotropy in Nematic Liquid Crystals**

Charles Parton-Barr[1] and Richard. J. Mandle*[1,2]

[1] School of Chemistry, University of Leeds, Leeds, UK, LS2 9JT
[2] School of Physics and Astronomy, University of Leeds, Leeds, UK, LS2 9JT

*r.mandle@leeds.ac.uk

Contents:

1) Graph model architectures
2) Graph model hyperparameters

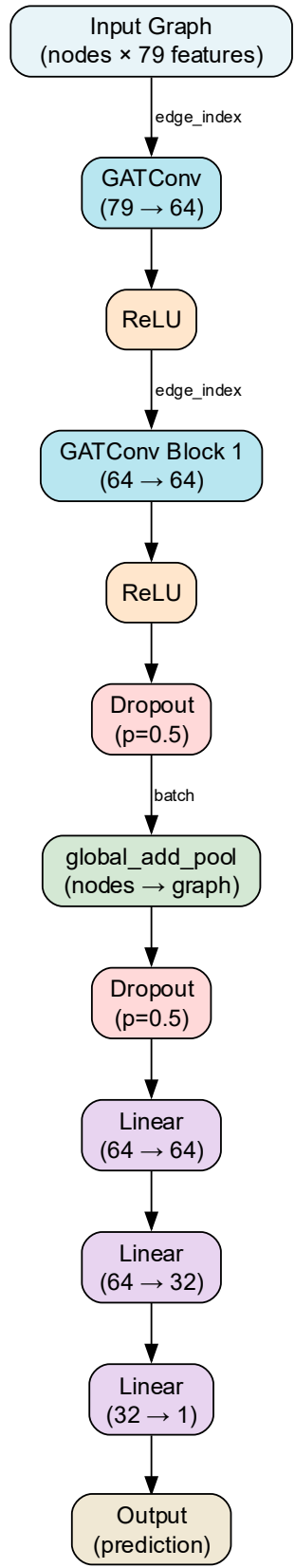

**Figure S1.** Schematic representation of the Predictor model architecture with example hyperparameters

**GATPredictor2**

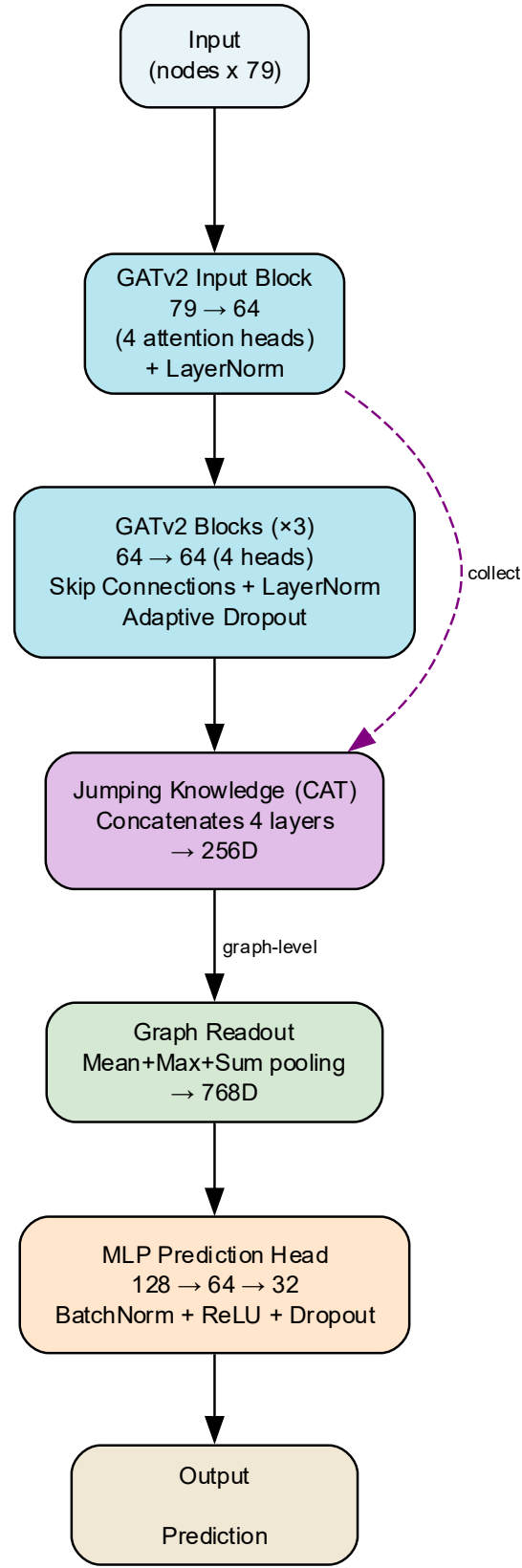

**Figure S2.** Schematic representation of the EnhancedPredictor model architecture with example hyperparameters. The architecture has been simplified for layout purpose

**GINPredictor**

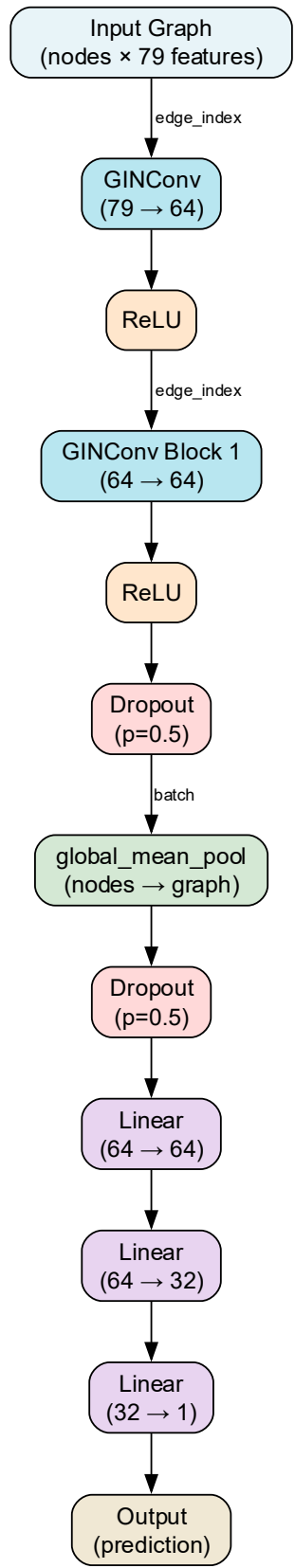

**Figure S3.** Schematic representation of the GINPredictor model architecture with example hyperparameters

**GCNPredictor**

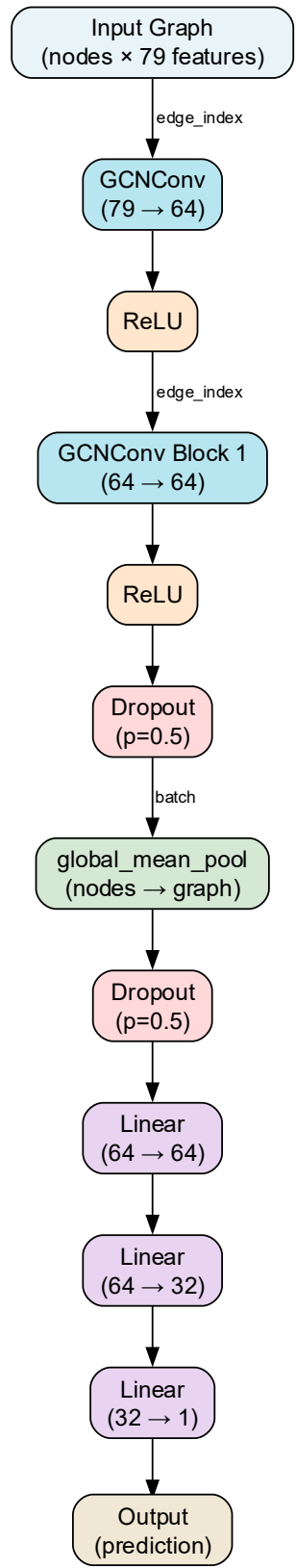

**Figure S4.** Schematic representation of the GCNpredictor model architecture with example hyperparameters

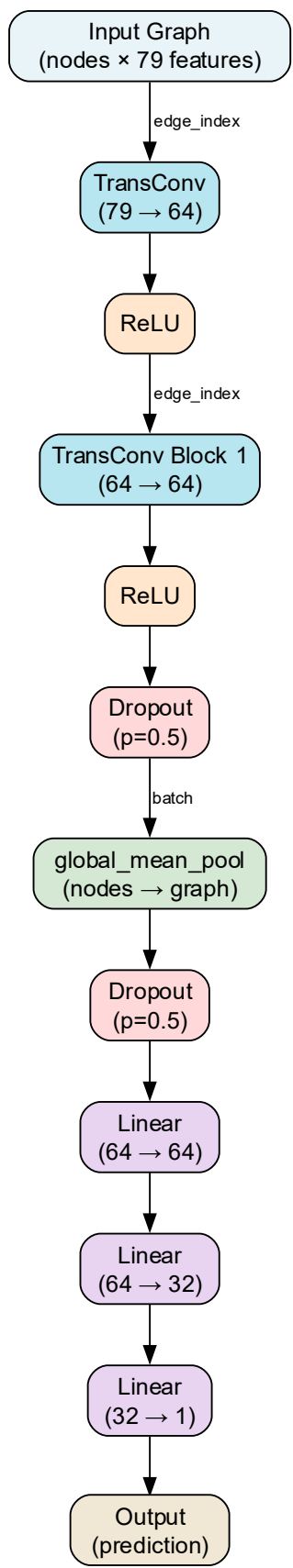

**Figure S5.** Schematic representation of the TransformerConv predictor model architecture with example hyperparameters

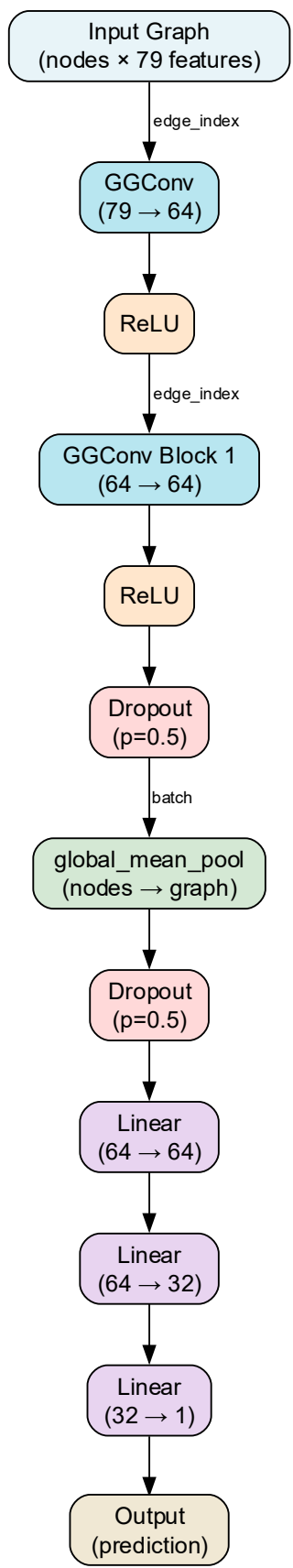

**Figure S6.** Schematic representation of the Gated Graph predictor model architecture with example hyperparameters

**Model hyperparameters**

| Model | Hyperparameters |
|---|---|
| GATPredictor | Model type: Predictor<br>Convolutional blocks: 3<br>Dropout: 0.15<br>Convolutional layer hidden dimension: 512<br>Learning rate: 0.0005<br>Batch size: 64<br>Epochs: 50<br>k-fold splits: 5 |
| GATPredictor2 | Model type: EnhancedPredictor<br>Convolutional blocks: 3<br>Dropout: 0.15<br>Convolutional layer hidden dimension: 512<br>Learning rate: 0.005<br>Batch size: 64<br>Epochs: 50<br>k-fold splits: 5 |
| GCNPredictor | Model type: GCNPredictor<br>Convolutional blocks: 3<br>Dropout: 0.15<br>Convolutional layer hidden dimension: 512<br>Learning rate: 0.0005<br>Batch size: 64<br>Epochs: 100<br>k-fold splits: 5 |
| GINPredictor | Model type: GINPredictor<br>Convolutional blocks: 3<br>Dropout: 0.15<br>Convolutional layer hidden dimension: 512<br>Learning rate: 0.001<br>Batch size: 64<br>Epochs: 100<br>k-fold splits: 5 |
| GatedGraphPredictor | Model type: GatedGraphPredictor<br>Convolutional blocks: 3<br>Dropout: 0.15<br>Convolutional layer hidden dimension: 256<br>Learning rate: 0.001<br>Batch size: 64<br>Epochs: 100<br>k-fold splits: 5 |
| TransformerPredictor | Model type: TransformerPredictor<br>Convolutional blocks: 3<br>Dropout: 0.15<br>Convolutional layer hidden dimension: 256<br>Learning rate: 0.0005<br>Batch size: 64<br>Epochs: 100<br>k-fold splits: 5 |

**Table S2:** Model hyperparameters.

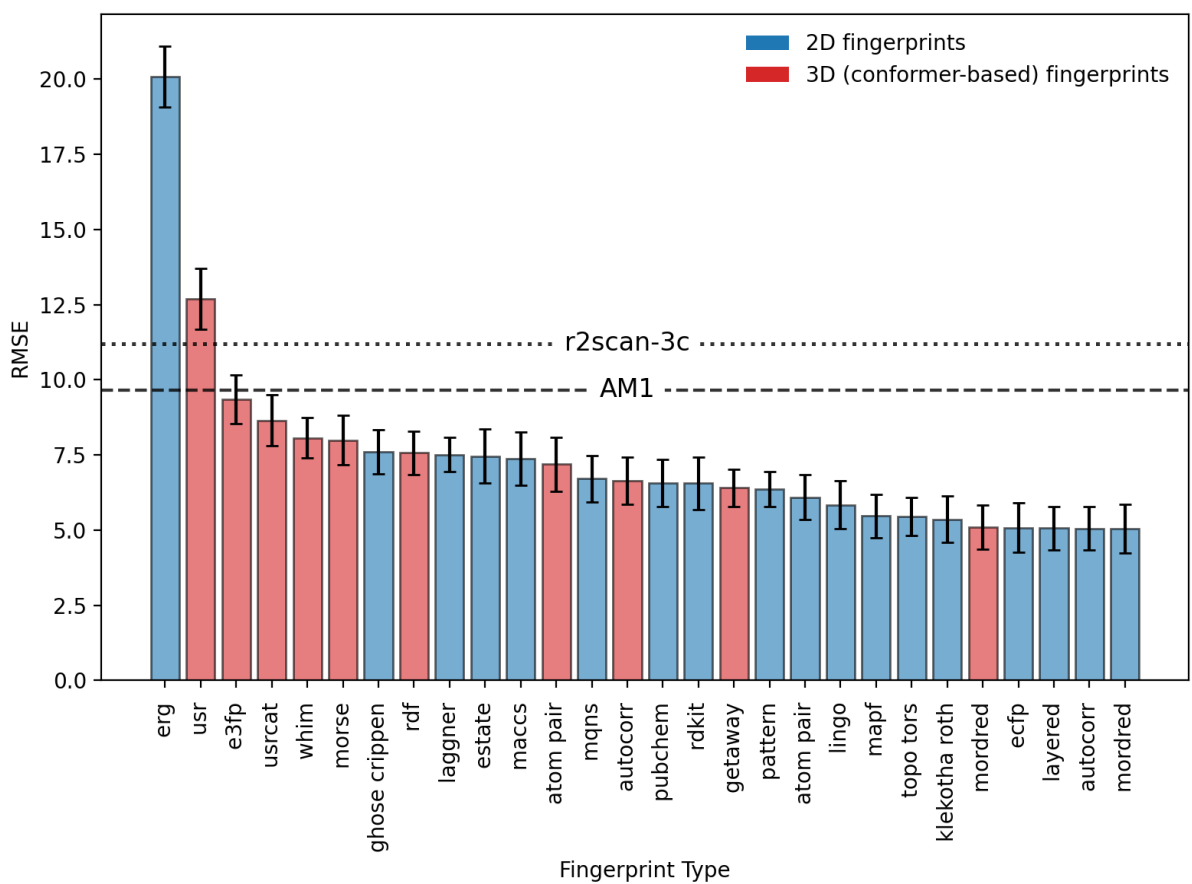

**Figure S7.** Performance of selected fingerprint based supervised machine-learning models in predicting dielectric anisotropy ($\Delta\epsilon$). Bars show the root-mean-square error (RMSE) on the independent hold out test set for each model type; error bars show the standard deviation of the RMSE across cross-validation folds, arising from the repeated data partitioning during model training. Horizontal dashed lines indicate the RMSE obtained for the same dataset when computing using the Maier-Meier relations with molecular properties as computed with the AM1 semi-empirical and r2scan-3c composite methods (*vide supra*). This Figure is a counterpart to Figure 4 in the main manuscript text, and showcases the RMSE of less performant networks.

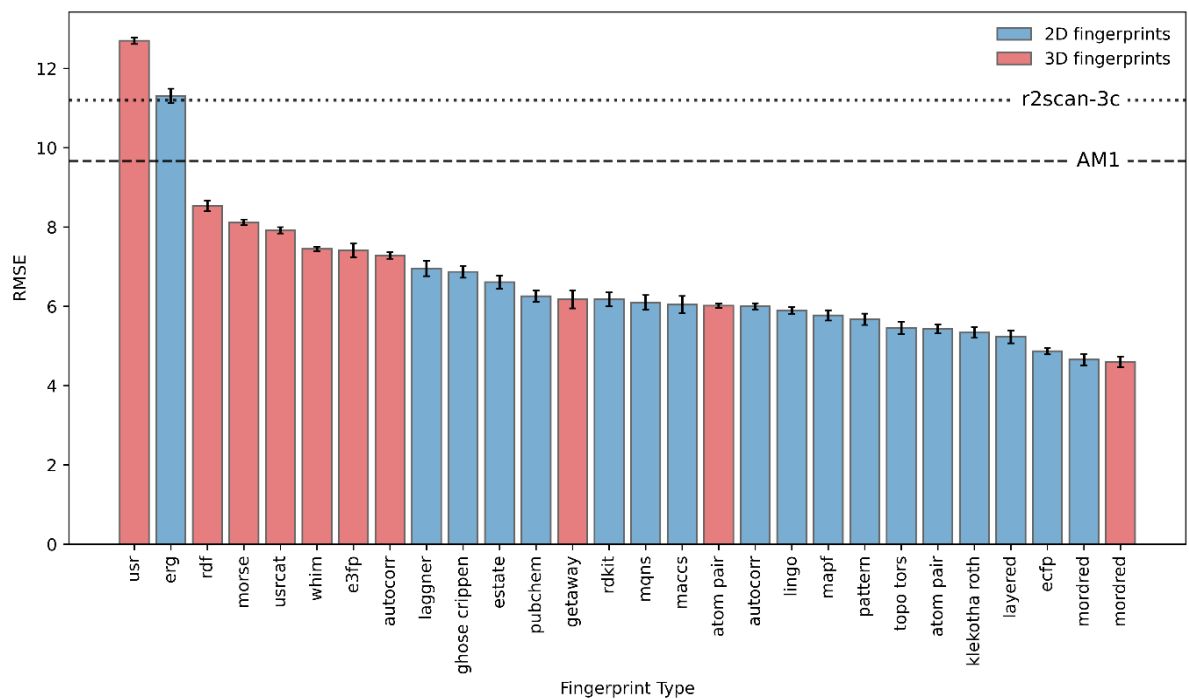

**Figure S8.** Performance of selected XGBoos regressor based supervised machine-learning models utilising molecular fingerprints in predicting dielectric anisotropy (Δϵ). Bars show the root-mean-square error (RMSE) on the independent hold out test set for each model type; error bars show the standard deviation of the RMSE across cross-validation folds, arising from the repeated data partitioning during model training. Horizontal dashed lines indicate the RMSE obtained for the same dataset when computing using the Maier-Meier relations with molecular properties as computed with the AM1 semi-empirical and r2scan-3c composite methods (*vide supra*). This Figure is a counterpart to Figure 4 in the main manuscript text, and showcases the RMSE of less performant XGBoost networks.

**Dataset Structure**

Below, in Table S2, we give selected examples from the dataset used in this work:

| Int. ID | Δε | smiles | REF |
|---|---|---|---|
| M_PK_1 | 9.7 | CCCC0CCC(C1CCC(c2cc(F)c(F)c(F)c2)CC1)CC0 | 10.1002/ejoc.200800149 |
| A_DOH_1 | -4.71 | CCC[C@H]1CC[C@H]([C@H]([C@@H](COC2=CC=C(CCC)C=C2)F)F)CC1 | 10.1016/j.tet.2014.05.027 |
| C_HY_13 | 11 | FCCCC0CCC(c1cc(F)c(c2cc(F)c(OC(F)(F)C(F)(F))cc2)c(F)c1)CC0 | US6248260B1 |
| J_YG_34 | 24.5 | CCCC0CCC(C1CCC(C(F)(F)Oc2cc(F)c(O/C=C\C(F)(F))c(F)c2)CC1)CC0 | US20150060732A1 |

**Table S2.** Structure of the dataset used in this work. **Int. ID** = An internal ID; **Δε** = reported dielectric anisotropy; **Phase Seq.** = Reported transition temperatures °C; **smiles** =

molecular structure as a smiles string; **REF** = reference; either a DOI number for journal articles or a patent number.

Discussing the dataset briefly, we note a few important considerations. Often brackets are used to denote monotropic phase transitions in the literature, however this is **entirely superfluous for most cases** as the absolute values of transitions indicate monotropic or enantiotropic phase transitions perfectly clearly. Although not used here, when parsing transitions we simply treat the melt (K XXX °C) as a special case, after which other transitions follow. The prediction of phase type and transition temperature from molecular structure will be the subject of a forthcoming paper. [1]

Given that different smiles strings can be used to describe the same molecule, it is especially important to perform duplicate checking on higher level objects after formal canonicalization (e.g. using rdkit.Chem.MolFromSmiles to obtain a *mol* object). Also, the use of [C@H] and [C@@H] to construct the proper stereochemistry around unsaturated ring systems is mandatory, as is the use of "/" and "\" to denote alkene stereochemistry.

Python code for reading the dataset, training models, generating citation data, performing dataset analysis and so on, is given at the Github page. This also contains a smaller dataset for testing purposes. [2]

**Supplementary references**

1. C. Parton-Barr, R. J. Mandle *et al*, *Forthcoming*.
2. C. Parton-Barr, R. J. Mandle, https://github.com/RichardMandle/ML_dielectric